\documentclass[10pt,conference]{IEEEtran}
\usepackage{bm}
\usepackage{amsmath}
\usepackage{amsfonts}

\newcommand{\R}{{\mathbb{R}}}

\DeclareMathOperator{\Cone}{Cone}
\DeclareMathOperator{\conv}{conv}
\DeclareMathOperator{\Exp}{Exp}
\DeclareMathOperator{\Cl}{cl}

\newcommand{\ve}[1]{\mathbf{#1}}

\newcommand{\beq}{\begin{equation}}
\newcommand{\eeq}{\end{equation}}
\newcommand{\beqa}{\begin{eqnarray}}
\newcommand{\eeqa}{\end{eqnarray}}
\def\cl{{\cal L}}

\newcommand{\ket}[1]{| #1 \rangle}
\newcommand{\bra}[1]{\langle #1 |}

\def\qed{$\Box$}

\def\half{{\frac{1}{2}}}
\def \qed {\hfill $\Box$\vspace{0mm}}
\newcommand{\eps}{\varepsilon}

\begin{document}

\title{Nonclassicality without entanglement enables bit commitment}


\author{
\authorblockN{Howard Barnum} 
\authorblockA{CCS-3: Information Sciences, and Quantum Institute \\
Los Alamos National Laboratory \\ Los Alamos, NM USA \\
Email: barnum@lanl.gov}
\and
\authorblockN{Oscar  C.O. Dahlsten} 
\authorblockA{Institute for Theoretical Physics \\ ETH Z\"urich,
Switzerland\\
Email: dahlsten@phys.etz.ch}
\and
\authorblockN{Matthew Leifer} 
\authorblockA{Perimeter Institute for Theoretical Physics \\ 
and  Institute for Quantum Computing \\ University of Waterloo \\Waterloo, Ontario, Canada \\
Email: matt@mattleifer.info}
\and 
\authorblockN{Ben Toner} 
\authorblockA{Centrum voor Wiskunde en Informatica \\ Amsterdam, The Netherlands \\
Email: Ben.Toner@cwi.nl} 
}
\maketitle
\begin{abstract}
We investigate the existence of secure bit commitment protocols in the
convex framework for probabilistic theories.  The framework makes only
minimal assumptions, and can be used to formalize quantum theory,
classical probability theory, and a host of other possibilities.  We
prove that in all such theories that are locally non-classical but do not
have entanglement, there exists a bit commitment protocol that is
exponentially secure in the number of systems used.
\end{abstract}
\section{Introduction}
In the 1984 paper \cite{Bennett84a} in which they introduced
information-theoretically
secure quantum key distribution, Bennett and
Brassard also considered the possibility of information-theoretically
secure bit commitment.  Bit commitment is a basic primitive in
classical cryptography, to which many
practically important cryptographic tasks, such as secure function evaluation, can be reduced.  In a bit commitment protocol, one
party, usually called Alice, performs some act that is supposed to
irrefutably convince another party, Bob, that she
has irrevocably committed to a value, 0 or 1, of a
bit, without leaking any information about the value of the
bit to Bob.  Later she can perform
another act that reveals the value of the bit to Bob and enables him to perform some test that
may be necessary for him to verify that she was indeed committed.    
Classically, bit commitment can be achieved with \emph{computational} security,  but not with information-theoretic security.

Bennett and Brassard showed that the bit commitment scheme they
considered could be defeated by the use of entangled
states.  Attempts were made \cite{BCJL93a}  to construct secure bit
commitment protocols, but Lo and Chau \cite{Lo97a}, and
independently Mayers \cite{Mayers97a}, showed that an entangled attack akin to Bennett
and Brassard's defeats all quantum bit commitment protocols, and
there is now a 
solid consensus that this does indeed cover all reasonable schemes and attacks
\cite{DAriano2007a}.

Soon after this development, 
Brassard \cite{Brassard:2005a} and 
Fuchs \cite{Fuchs:2003} 
asked whether the
impossibility of bit commitment might be a manifestation of a deep
information-theoretic property of quantum mechanics, fit for a crucial
role in an information-theoretic characterization, or reconstruction,
of the 
formalism of quantum theory.  Such a
reconstruction, at its most ambitious, is envisioned as 
similar to Einstein's reconstruction of the dynamics and kinetics of
macroscopic bodies on
the basis of simple principles with clear operational meanings 
and 
experimental consequences.  
As argued in (for example)   
\cite{Clifton:2003, Grinbaum:2007a, Fuchs:2003}, 
such a reconstruction 
could lend force to
the view that the foundations of quantum mechanics are properly couched in terms of
information, a view which has received increasing attention with the rise of
quantum information science.  Short of this ambitious
goal, there are still strong reasons to pursue an informational
characterization of quantum mechanics.  It should lead to a
principled understanding of the features of quantum mechanics that
account for its better-than-classical information processing power.  Such an 
understanding could help guide the search for new
algorithms and protocols, both positively, by providing conceptual
tools to exploit in a variety of settings, and negatively by
identifying information-processing tasks requiring properties that
quantum mechanics 
lacks.

Brassard and Fuchs' conjecture was that the impossibility of bit
commitment might, in conjunction with the possibility of secure secret
key distribution and the impossibility of instantaneous signaling
between distinct physical systems, suffice to characterize quantum
theory.  Clifton, Bub, and Halvorson proved a result (the CBH theorem)
\cite{Clifton:2003}, close to this conjecture in the framework of
$C^*$-algebraic theories.  They demonstrated the existence of a
protocol related to the no-bit commitment theorem, but weaker, between
two ``local'' algebras, whenever the local algebras are not
commutative (not classical) and there are entangled states between the
algebras.  However, in finite dimensions, $C^*$-algebraic theories are
essentially quantum mechanics with superselection rules, so in our
view, a much broader framework is desirable.  Further evidence for
this view is Halvorson's demonstration \cite{Halvorson04} that
no-bit-commitment 
follows from 
no-signaling and
no-cloning 
within the $C^*$-algebraic framework.  To obtain the
most illuminating characterization of quantum mechanics in terms of
information processing, one should work in a framework wide enough to
include not only quantum and classical mechanics, but also a wide
variety of other theories that can serve as foils to them;
the $C^*$-algebraic framework is too restrictive.

It is therefore an open question whether non-classical theories
without entanglement are ruled out by demanding the impossibility of
secure bit commitment, in some appropriately broad framework.  In this
paper, we answer that question in the affirmative.  We work in a
framework that allows for a wide range of probabilistic theories,
including not only quantum and classical theories, but also theories
of Popescu-Rohrlich, or nonlocal, boxes \cite{Popescu:1994,
Barrett:2005a} that allow nonlocality stronger than that in quantum
mechanics, as well as many other types of theory.  For any
nonclassical theory within the framework that does not permit
entanglement between systems, we construct a bit-commitment protocol
that is exponentially secure in the number of systems used.

 We proceed as follows. First the framework of generalized
probabilistic theories is introduced and our bit-commitment protocol is
defined. We then prove that such a protocol
always exists in a non-classical theory.  Next, we prove it
to be exponentially secure in all theories that don't allow
entanglement.  Finally we give a summary and discussion.


\section{The Framework}
The framework is that of {\it convex operational} or {\em generalized
  probabilistic} theories, for which no-cloning and no-broadcasting
  theorems were proved in \cite{BBLW20072, BBLW2007}, to
  which we refer for further background.  The set of
  \emph{normalized states} of a system is a compact convex set $\Omega
  \subseteq\R^d$. Embed $\Omega$ in $\R^{d+1}$, 
avoiding the origin, 
and let $\Cone(\Omega)$ be the set of
  linear combinations of elements of $\Omega$ with nonnegative
  coefficients---the convex cone of
  \emph{unnormalized} states.  Its \emph{dual cone}, 
$\Cone(\Omega)^*$,
  consists of those linear functionals from $\R^{d+1}$
  to $\R$ that are nonnegative on
  $\Cone(\Omega)$. Measurement outcomes are represented as {\em effects}:
functionals  
$e \in \Cone(\Omega)^*$ satisfying $e(\omega) \le 1$ for all $\omega \in \Omega$.  $e(\omega)$ is
the 
probability of outcome $e$ for a system prepared in state $\omega$.  
Equivalently, effects are 
elements of the interval $[0,u]$ in the dual
  cone, whose endpoints are the zero functional and the unit
  functional $u$ that gives $1$ on all normalized states.
{\em Measurements} are sets $\{e_i\}$ of effects with $\sum_i e_i = u$ 
(i.e. $\forall \omega \in \Omega, \sum_i e_i(\omega) = 1$).

For two state spaces, $\Omega_A$ and $\Omega_B$, a spectrum of
possible ``tensor products'' is identified---these are candidates for
describing a composite system built from subsystems with state spaces
$\Omega_A$ and $\Omega_B$.  In this work we need only one:

\emph{Definition}: The \emph{minimal tensor product} $\Omega_A \otimes
 \Omega_B$ is the convex hull of the set of product states $(\omega_A,
 \omega_B) \in \Omega_A \times \Omega_B$.

This generalizes the quantum-mechanical construction of the
unentangled or \emph{separable} density matrices.  The general
framework requires only that a tensor product be convex, contain the
minimal tensor product, and be contained in what's known as the
\emph{maximal tensor product}, of less interest here.

To describe quantum theory in this framework, $\Omega$ is chosen to be
isomorphic to the set of density operators on a Hilbert space and
$\Cone(\Omega)$ is the set of positive operators.  The quantum tensor
product lies strictly between the minimal and maximal tensor
products. In classical theory, $\Omega$ is a simplex of probability
distributions, i.e. the convex hull of $d+1$ linearly independent
points in $\R^{d+1}$, and the maximal and minimal tensor products
coincide so there is no choice.  Classical theories are,
equivalently, characterized by the property that any state in $\Omega$
has a unique convex decomposition into pure (extremal) elements.

It is important to specify the dynamics of theories in this framework,
because this specifies what Alice and Bob can do to their systems.  In
this framework, dynamics are {\em positive} linear maps $\cl: \R^{d+1}
\rightarrow \R^{d+1}$, i.e. ones that take $\Cone(\Omega)$ to itself.
Thus they take (not-necessarily-normalized) states to states.
Further, they must be {\em norm-nonincreasing}: for all states $\omega
\in \Cone(\Omega)$, $u(\cl(\omega)) \le u(\omega)$; we use the term
{\em operation}, standard for the quantum case, to denote these.  The
map $e_{\cl}: \omega \mapsto u(\cl(\omega))$ is an element of $[0,u]$,
and is interpreted as an effect (measurement outcome) associated with
the dynamics $\cl$.  Thus for normalized $\omega$ and positive $\cl$,
$e_{\cl}(\omega)$ is interpreted as the probability with which the
state undergoes $\cl$.  When $e_{\cl} = u$, the map is {\em
norm-preserving}; it is an {\em unconditional} dynamics not associated
with obtaining a particular measurement outcome.


Early work on cryptography using stronger-than-quantum
nonlocal correlations, including \cite{BCUWW2005a} and
\cite{ShortGisinPopescu2005a} where {\em entangled} correlations 
enabled bit commitment, did not situate these
correlations in a unified framework describing dynamics, measurement,
and state preparation such as the one we use here.

The assumptions embodied in this framework \cite{BBLW2007} are fairly
minimal.  Two are substantive: first, the ``local observability''
assumption effectively states that there are no ``intrinsically
nonlocal'' degrees of freedom that cannot be determined by making
repeated local measurements on the subsystems of identically prepared
systems.  Second,
a ``no-signaling'' constraint, which it is reasonable
to take as the definition of what we mean by an independent subsystem.

Our protocol uses the fact that any nonclassical state-space
contains states that have more than one distinct convex decomposition into pure states.
Alice encodes which bit she has committed to as a choice of one out of two such 
decompositions.  The security
analysis we give requires that the two sets of pure states used in the decompositions 
be {\em disjoint}, and that all the states be {\em exposed}, but this 
can be achieved in any nonclassical state space.  A state is {\em exposed} if there is 
a measurement outcome whose probability is $1$ in that state, and strictly less than $1$
on any other state---an outcome that can be guaranteed by that state, and only by that 
state.    We call such an effect the {\em distinguishing effect} for the state in question.  
It is immediate from the definitions that exposed states are pure.  

We write $\Cl(S)$, $\conv(S)$, and $\Exp(S)$ for 
the topological closure, convex hull, and set of exposed points of a set $S$.

\section{The Protocol}
 Let a system have a non-simplicial, convex, compact state space
$\Omega$ of dimension $d$.  
The  protocol uses a state $\mu$ that has two distinct convex decompositions 
$\{(p^0_i, \mu^0_i)\}, \{(p^1_j, \mu^1_j)\}$
into
finite disjoint sets  of exposed states,  that is, 
\begin{align}
\label{2ExpDecomp}
\mu = \sum_{i = 1}^{ N^0} p_i^0 \mu^0_i = \sum_{j=1}^{ N^1} p_j^1 \mu^1_j\;.
\end{align}

In the honest protocol, Alice first decides on a bit $b \in \{0,1\}$
to commit to.  She then draws $n$ independent samples from the probability distribution $(p^b_1, p^b_2,\ldots, p^b_{N^b})$, obtaining
a string $\ve{x} = (x_1,x_2,\ldots,x_n )$.  She sends the
state $\mu^b_{\ve{x}} = \mu^b_{x_1} \otimes \mu^b_{x_2} \otimes \ldots
\otimes \mu^b_{x_n}$ to Bob.


In the reveal phase, she sends $b$ and $\ve{x}$ to Bob. 
Bob then measures each subsystem of the state Alice sent in the commit phase.  
On the $k$-th subsystem, he performs a measurement containing the distinguishing effect for $\mu^b_{x_k}$ and
aborts if the result is not the distinguishing effect. 
If he obtains the appropriate distinguishing effect for every subsystem, he accepts.

{\em Example of protocol:} If $\Omega$ is the 
state space of a qubit, we can transpose the one-qubit protocol of \cite{Bennett84a}
to our setting. $\Omega$ can be visualised as the Bloch sphere in
$\R^3$ with pure states on the surface and their mixtures
inside the sphere. Let $\mu$ be the center of the sphere, i.e. the
completely mixed state $\half I =\frac{1}{2}\ket{+}\bra{+}+\frac{1}{2}\ket{-}\bra{-}=\frac{1}{2}\ket{0}\bra{0}+\frac{1}{2}\ket{1}\bra{1}$, where $\ket{0}, \ket{1}$ is a basis and $\ket{\pm} = \frac{1}{\sqrt{2}} (\ket{0} \pm \ket{1})$.
Let $\mu^0_1=\ket{0}\bra{0}$, $\mu^0_2=\ket{1}\bra{1}$,
$\mu^1_1=\ket{+}\bra{+}$, $\mu^1_2=\ket{-}\bra{-}$ and
$p^b_i=\frac{1}{2}\, \forall i,b$. In the $n=1$ case, if Alice decides to
commit to $b=0$ for example, she would send Bob either $\ket{0}$ or $\ket{1}$, each with
probability $\frac{1}{2}$.  Say she sends $\ket{0}$.  To reveal
she tells him ``$b=0$" and that she sent $\ket{0}$. Bob
would then measure in the $\ket{0}$, $\ket{1}$ basis,
find $\ket{0}$ and accept.  In \cite{Bennett84a}, Bennett and Brassard considered this $n=1$
protocol and showed it was completely nonbinding 
through an entangled attack.

\section{Existence of the Protocol}
The existence of the protocol just described  in any non-classical theory
follows from:

{\em Theorem 1:} Every nonsimplicial convex compact set $\Omega$ of
dimension $d$ contains a state $\mu$ with two convex decompositions
into disjoint sets of exposed states, whose total cardinality is less
than $d+2$.

The theorem follows from two lemmas.

{\em Lemma 1: }
Let $\Omega$ be a non-simplicial compact convex set of dimension $d$. 
Then the convex hull of any $d+2$ pure states in $\Omega$ contains 
a state $\mu$ which has two convex decompositions, 
\beq
\label{2Decomp}
\mu = \sum_{i = 1}^{ N^0} p_i^0 \mu^0_i = \sum_{j=1}^{ N^1} p_j^1 \mu^1_j,
\eeq
into disjoint sets of pure states, with $N^0 + N^1 \leq d+2$.



{\em Proof:}  
Let $\Gamma := \{\mu_1,..., \mu_{d+2}\}$ be an arbitrary  set of $d+2$ pure states.  
Then conv$(\Gamma)$ is 
  non-simplicial because $\Omega$ has dimension $d$. 
Choose a state
  $\omega$ with two different convex decompositions $\{(p^0_i, \mu^0_i), i = 1,.., N^0\}$ and 
$\{(p^1_j, \mu^1_j), j \in 1,...,N^1\}$ into elements of 
$\Gamma$ so that $N^0+N^1$ is
  minimal.  The sets $\{\mu^0_1, \ldots, \mu^0_{N^0}\}$ and
  $\{\mu^1_1, \ldots, \mu^1_{N^1}\}$ are then disjoint.  For if they had 
  a state in common, say (reindexing if necessary) $\mu^0_1 =\mu^1_1$, then the (unnormalized) state
  $\omega' := \omega - \min_b(p_1^b) \mu^b_1$ would also have two different
  convex decompositions, contradicting minimality.\qed


To show there are $d+2$ {\em exposed} states we'll use the following special case of Theorem 18.7 of
\cite{Rockafellar70a}.


{\em Theorem 2: }
A compact convex set $\Omega \subseteq \R^d$ is the closure of the convex hull
of its exposed points, i.e. $\Omega = {\rm \Cl}({\conv}({\Exp}(\Omega)))$.

{\em Lemma 2:}
A nonsimplicial convex compact set $\Omega$ of dimension $d$ has at least $d+2$ exposed points.
%

{\em Proof:}
By Theorem 2, 
the closure of the convex hull of $\Exp(C)$ is equal to $C$, and therefore 
$\Cl(\Cone(\Exp(C))) = \Cone(C)$.  
Taking the closure of a convex subset (compact or not) 
of $\R^{n}$ can't increase the
dimension of the subspace it spans, so the linear span of $\Exp(C)$ must be $\R^{d+1}$, 
and we may pick a linearly independent subset of $\Exp(C)$, consisting of $d+1$ 
exposed points.  There must be an 
exposed point not in the convex hull of these $d+1$ points, for if not the convex hull of
the exposed extreme points of $C$ would be a simplex, whence, using Theorem 
2 and the fact that a finite-dimensional
simplex is closed, $C$ itself would be a simplex. \qed

Since exposed states are pure, Lemmas 1 and 2 immediately imply Theorem 1.
\\

\section{Security of the Protocol}
We adapt our security definition from Ref.~\cite{buhrman:_possib_impos_cheat_sensit_quant}, simplifying to the setting where there is no communication from Bob to Alice. We start with the formal definition:

\emph{Definition}: Let $\eps \geq 0$. We say that a bit commitment
protocol with one-way communication is $\eps$-\emph{secure} if it has the
following properties:
\begin{itemize}
\item ($\eps$-\emph{soundness}) Assume that both parties are
  honest. Then the probability that Bob aborts is at most $\eps$ and,
  if he does not abort, then after the reveal phase he learns the bit
  $b$ that Alice committed to.

\item ($\eps$-\emph{hiding}) Assume that Alice is honest.  Then for all 
cheating strategies of Bob aiming to guess the committment before the 
reveal phase. $q_0 + q_1 \leq 1 + \epsilon$, where $q_b$ is the probability
that Bob guesses correctly given that Alice committed $b$.


\item ($\eps$-\emph{binding}) Assume that Bob is honest. Then for all
commitments of Alice, $p_0 + p_1 \leq 1 + \eps$, where $p_b$ is the
maximum probability that Alice 
successfully 
reveals $b$.
\end{itemize}
If any of the above hold for $\eps=0$, we say that the protocol satisfies that property \emph{perfectly}.
 
Our protocol is perfectly \emph{sound} because if Alice is honest, the
distinguishing measurements that Bob makes based on Alice's claim give
the correct answers with probability $1$.  In general, one would
consider the probability of \emph{either} honest participant accusing
the other of cheating, but in a one-way protocol, there is no
provision for Alice to abort.

 The protocol is also perfectly \emph{hiding}---there is no way for Bob to obtain
information about the bit $b$ during the post-commit, pre-revelation phase, 
as the state $\mu^{\otimes n}$ that (honest) Alice sent is independent of $b$.

The nontrivial part of the security analysis is to show that the protocol is $\eps$-\emph{binding}---to show
that Alice can't cheat by choosing which bit to 
reveal after she is supposed to be committed to one or the other. In an ideal bit commitment
protocol, Alice could use randomness 
to commit to $0$ with probability $p_0$ and
$1$ with probabilities $p_1=1-p_0$, so she can achieve any pair $p_0, p_1$ 
in the definition such that $p_0 + p_1 = 1$. Our protocol only allows her to do a little better. For example, if she wants to be able to reveal $0$ with
probability $1$, then the probability that she can reveal $1$ is at most $\eps$. Although it is suitable for present purposes, we note that our definition of $\eps$-binding is too weak to establish composable security~\cite{damgard07:_tight_high_order_entrop_quant}.

We'll need a lemma about measurements.

{\em Lemma 3:}  
Suppose two exposed states $\mu \neq \nu$ have distinguishing effects $a$ and $b$. Let 
  \begin{align}
    f(\mu, \nu) := \sup_{\omega \in \Omega}\left( a(\omega) + b(\omega) \right).
  \end{align}
Then $1 \leq f(\mu, \nu) < 2$.

{\em Proof: } 
 For the upper bound, the function $a + b$ is linear and the set $\Omega$ is convex and
  compact, so the supremum of $a +b$ is achieved on a pure state
  $\omega'$. Suppose $a(\omega') + b(\omega') = 2$. Then we must have
  $a(\omega') = 1$ and $b(\omega') =1$, which implies $\omega' = \mu =
  \nu$, a contradiction.  The lower bound follows from considering $\omega = \mu$.\qed

Now define 
$  
\delta := \min_{1\leq i \leq N^0, 1 \leq j \leq N^1} \left(2 - f(\mu^0_i, \mu^1_j)\right),
$
where $\mu^0_i, \mu^1_j$ run over the states used in the
protocol. Note that $\delta < 1$, since at least one pair of states $\mu^0_1, \mu^1_j$ is not perfectly distinguishable.  This quantity $\delta$ will control the number $n$ of systems
we need to use to achieve $\eps$-security.

The proof also uses the following description of an optimal set of strategies
for a cheating Alice.

{\em Lemma 4:}
An optimal strategy for Alice is as follows:  
she tosses some coins and generates randomness $\lambda$ with
probability weight $p(\lambda)$.  She then
prepares an arbitrary string of pure states $\omega_1^\lambda \otimes
\omega_2^\lambda \otimes \ldots \otimes \omega_n^\lambda$. 
 She
sends them to Bob.  In the reveal phase, she can send an arbitrary bit
$b$ and an arbitrary ``claim sequence'' $\ve{x}^{\lambda, b}$, that depends on the
bit she wants to claim and the randomness.  

The state claim $\ve{x}^{\lambda, b}$, which is classical information,
is encoded in perfectly distinguishable states of some systems $\Gamma$ in the
theory; 
it is easily shown that doing otherwise
can't help Alice.        

{\em Proof of Lemma 4:}  A general cheating strategy for Alice is to prepare
an arbitrary state in $\Upsilon \otimes \Gamma^{\otimes  n} \otimes \Omega^{\otimes  n}$ 
(where $\Upsilon$ is some state space in the theory), and then do a $b$-dependent 
positive map $\cl^b$ on $\Upsilon \otimes \Gamma^{\otimes  n}$ just before sending $\Gamma^{\otimes  n}$ to Bob, in an attempt to reveal $b$.  
Letting $r^l$ be probabilities, $\tau^l \in \Upsilon$, $\gamma^l_k \in 
\Gamma$, $\omega^l_k \in \Omega$, the state before revelation is:
\beq
\sum_{l} r^{l}
\tau^l \otimes \gamma_1^l \otimes \cdots \gamma_{ n}^l \otimes \omega_1^l \otimes
\cdots \otimes
\omega_{ n}^l\;.
\eeq
After Alice attempts to reveal $b$ it is:
\beq
\psi^b := 
\sum_{l} r^{l}
\cl^b(\tau^l \otimes \gamma_1^1 \otimes \cdots \otimes \gamma_{ n}^l) \otimes 
\omega_1^l \otimes \cdots \otimes \omega_{ n}^l\;.
\eeq
Let $\phi^{lb} \equiv \sum_m t^{lmb} \gamma_1^{lmb} \otimes \cdots \otimes \gamma_{ n}^{lmb}$ 
be the marginal state on $\Gamma^{\otimes  n}$ induced by the state
$\cl^b(\tau^l \otimes \gamma^l_1 \otimes \cdots \otimes \gamma^l_{ n})$.  
Then the state of $\Gamma^{\otimes  n} \otimes \Omega^{\otimes  n}$ is
\beq
\chi^b := \sum_{l} r^{l}
\sum_m t^{lmb} \gamma_1^{lmb} \otimes \cdots \otimes \gamma_{ n}^{lmb}
 \otimes 
\omega_1^l \otimes \cdots \otimes \omega_{ n}^l\;.
\eeq
Bob will subject each copy of $\Gamma$ to a standard measurement to 
read Alice's claim; the $k$-th system will yield a value $x$ with 
probability $p^{lm}_k(x)$ determined by $\gamma^{lm}_k$.  Alice could
achieve the same result by sampling the distribution of measurement
results $p^{lm}_k$ Bob would obtain from $\gamma^{lm}_k$, perhaps
keeping a record $q$ of the result of sampling, and sending a {\em definite} string 
$x^{lmqb}_1 \otimes \cdots x_{ n}^{lmqb}$ for the claim, encoded as distinguishable states that will definitely give this string of outcomes.
Letting $\lambda$ stand for $lmq$, we see that an optimal strategy for Alice is as
described in the Lemma.  \qed

{\em Theorem 3: } Our bit commitment protocol is $\eps$-binding with $\eps = \left(1-\delta\right)^n$.

{\em Proof: } 
Let $a^b_i$ be the distinguishing effect
for $\mu^b_i$ ($b \in \{0,1\}, i \in \{1,...,N^b\}$). 
Define $q_k^b(\lambda) :=
a^b_{x^{\lambda,b}_k} (\omega_k^\lambda)$; this is the probability that
$\omega_k^\lambda$ passes the test Bob performs on the $k$-th system 
in the reveal phase when Alice tries to reveal $b$. 
Then
$ 
 q_k^0(\lambda) + q_k^1(\lambda) \leq 2 - \delta,
$
by our choice of $\delta$. 

Since Bob only accepts if he accepts the state $\omega^\lambda_k$ of each subsystem, we have:
\begin{align}
  p_0 + p_1 = \sum_\lambda p(\lambda) \left[ \prod_{k=1}^n q_k^0(\lambda) + \prod_{k=1}^n q_k^1(\lambda) \right].
\end{align}
By convexity, we can fix some best choice for the randomness $\lambda$ and drop the label.
An upper bound on 
\begin{align} \label{eq: objective function}
  p_0 + p_1 = \prod_{k=1}^n q_k^0 + \prod_{k=1}^n q_k^1,
\end{align}
is obtained by  maximizing it 
subject to $0 \leq q_k^0, q_k^1 \leq 1$ and $q_k^0 + q_k^1 \leq 2 - \delta$. 
We should saturate the second inequality, since adding to $q_k^0$ or $q_k^1$ can only increase the right-hand side of 
Eq.~(\ref{eq: objective function}).  Now let 
$
Q_{\hat k}^b := \prod_{\substack{k=1..n\\k \neq \hat k} } q_k^b\; ,
$
so that 
$
  p_0 + p_1 = Q_{\hat k}^0 q_{\hat k}^0 + Q_{\hat k}^1 q_{\hat k}^1.
$
Since this expression is affine in $q_{\hat k}^0$, it's clear that if
$Q_{\hat k}^0 > Q_{\hat k}^1$, we should take $q_{\hat k}^0 = 1$ and
$q_{\hat k }^1 = 1-\delta$, and vice versa if $Q_{\hat k}^1 > Q_{\hat
k}^0$. If $Q_{\hat k}^1 = Q_{\hat k}^0$, then we can take either
$q_{\hat k}^0 = 1$ and $q_{\hat k }^1 = 1-\delta$ or use the
opposite assignment. 
Therefore,
\begin{align}
  p_0 + p_1 \leq \max_{m=0..\lfloor n/2\rfloor} (1-\delta)^m + (1-\delta)^{n-m}.
\end{align}

If $0< m < n/2$, then we can increase the sum by moving a $1-\delta$
term from $(1-\delta)^m$ to $(1-\delta)^{n-m}$, from which it
follows that
\begin{align*}
p_0 + p_1 \leq \begin{cases}1 + (1-\delta)^n & \text{if $n$ is odd;}\\
\max(1 + (1-\delta)^n, 2 (1-\delta)^{n/2}) & \text{if $n$ is even.}
\end{cases}
\end{align*}
For even $n$, note that $1 + (1-\delta)^n - 2 (1-\delta)^{n/2} =
(1 - (1-\delta)^{n/2})^2 \geq 0$, so the maximum is always achieved
by the first term. This proves the theorem.\qed


\section{Related Work}
Winter, Nascimento, and Imai 
\cite{WNI2003a} 
found 
the optimal rate at which a discrete memoryless classical channel from
Alice to Bob can be used to commit bits.  Because the set of
achievable output distributions may be a nonsimplicial compact convex
body $\Omega$, and the channel allows Alice to prepare any
distribution of products of states in this convex body, their setting
has similarities with ours.  But it permits only a fixed output
measurement 
whereas ours permits any measurement of effects in
the cone dual to this convex body.  Our setting also differs
by permitting unentangled nonclassical processing
by Alice and Bob.  Also, the discreteness of the classical channel implies
that the set of possible output distributions for the
channel is a polytope, whereas in our theories $\Omega$ can be an
arbitrary compact convex body.  Finally, we do not calculate rates, but 
demonstrate exponentially secure commitment of a single bit;
bounding the rate in our theories would be interesting, but
it is not obvious what good analogues of the bounding entropic expressions in
\cite{WNI2003a} would be.


Wolf and Wullschleger (WW) \cite{WolfWullschleger2005b} reach a
conclusion qualitatively similar to ours, that in a setting
more general than quantum theory, assumptions that rule out
entanglement can provide a secure protocol.  They have told us
that their result will be strengthened in \cite{SeverinWW08}.
\cite{WolfWullschleger2005b} assumes Alice and Bob have access to many
independent uses of the same trusted bipartite box-pair, initially
uncorrelated with anything else.  The boxes have binary inputs and
outputs, but WW state that extension to larger finite sets of inputs and
outputs is straightforward.  Under the very weak condition that one
party's conditional state depends on the other's input, they
provide a bit commitment protocol and a security proof.
Our setting is more general as it does 
not assume a trusted joint Alice-Bob state. 

\section{Conclusion and Discussion}

In \cite{BBLW20072, BBLW2007, Barrett:2005}, it was shown that the
no-broadcasting and no-cloning theorems, and the tradeoff between
information gain and state disturbance, are generic in non-classical
theories in our framework.  For the project of characterizing quantum mechanics 
this focuses attention on properties, like the
impossibility of bit commitment and the possibility of teleportation,
that may {\em not} be generically non-classical.



Within our framework, if one makes the plausible assumption that
an information-disturbance tradeoff (which is equivalent to
nonclassicality) allows secure key distribution, we may paraphrase the
Brassard-Fuchs conjecture as saying that the impossibility of bit
commitment characterizes quantum mechanics from among the nonclassical
theories in our framework.  We have shown that nonclassical theories
in which bit commitment is impossible must have entanglement, but in
contrast to the situation for the $C^*$-algebraic framework, in the
general framework that is very far from narrowing us down to quantum
theory.  An important open question, then, is what, if any,
sorts of theories in our framework that {\em do} have entanglement,
nevertheless permit bit commitment.

\section*{Acknowledgments}
Part of this work was completed at the workshop ``Operational
probabilistic theories as foils to quantum theory'', July 2-13 2007 at
the University of Cambridge, funded by The Foundational Questions
Institute (FQXi) and SECOQC.  At IQC, ML was supported in part by
MITACS and ORDCF.  ML and OD were supported in part by grant
RFP1-06-006 from FQXi.  Research at Perimeter Institute for
Theoretical Physics is supported in part by the Government of Canada
through NSERC and by the Province of Ontario through MRI. BT is
supported by the EU FP6-FET Integrated Project QAP CT-015848, NWO VICI
project 639-023-302, and the Dutch BSIK/BRICKS project. HB was
supported by the US Department of Energy through the LDRD program at
LANL.





%



\end{document}